\documentclass[12pt]{article}
\usepackage[esperanto,portuguese]{babel}
\usepackage{epsfig} 
\usepackage{icomma}

\newcommand{\dd}{\mathrm{d}}

\newcommand{\ppunu}[1]{#1} \newcommand{\ppdu}[1]{}
\ppdu{
\newlength{\pplw}\setlength{\pplw}{0.453\textwidth}
\newlength{\pprw}\setlength{\pprw}{0.507\textwidth}

\newcommand{\ppp}{\ParallelPar}
\newcommand{\ppn}{\noindent}              %para n�o fazer par�grafo
\newcommand{\ppl}[1]{\ParallelLText{\selectlanguage{esperanto}#1}}
\newcommand{\ppr}[1]{\ParallelRText{\selectlanguage{portuguese}#1}\ppp}
\newcommand{\ppln}[1]
{\ParallelLText{\ppn \selectlanguage{esperanto}#1}} %p/ nao parag
\newcommand{\pprn}[1]
{\ParallelRText{\ppn \selectlanguage{portuguese}#1}\ppp} %p/ nao paragrafo

\newcommand{\ppsection}[3][0ex]{\vspace{2em} 
\ppl{\section{#2} \vspace{#1}} \ppa \nopagebreak
\ppR{\section{#3}} \ppp \nopagebreak}

\newcommand{\bea}{\vspace{-1ex}\begin{eqnarray}}
\newcommand{\eea}{\end{eqnarray}}
}

\ppunu{
\newcommand{\ppl}[1]{\selectlanguage{esperanto}#1}
\newcommand{\ppln}[1]{\noindent \selectlanguage{esperanto}#1}
\newcommand{\ppr}[1]{\selectlanguage{portuguese}}
\newcommand{\pprn}[1]{\selectlanguage{portuguese}}
\newcommand{\ppsection}[3][0ex]{\section{#2}}

\newcommand{\bea}{\begin{eqnarray}}
\newcommand{\eea}{\end{eqnarray}}
}

\title{\bf Dopplera efiko de luma ebeno vidata per akcelata observanto
}
\author{F.M. Paiva \\ 
{\small Departamento de F\'\i sica, U.E. Humait\'a II, Col\'egio Pedro II} \\
{\small Rua Humait\'a 80, 22261-040  Rio de Janeiro-RJ, Brasil; fmpaiva@cbpf.br} 
\vspace{.7ex} \\
A.F.F. Teixeira \\
{\small Centro Brasileiro de Pesquisas F\'\i sicas} \\
{\small 22290-180 Rio de Janeiro-RJ, Brasil; teixeira@cbpf.br}} 

\begin{document}
%\special{papersize=210mm,297mm}
\selectlanguage{esperanto}
\maketitle 
\thispagestyle{empty}

\begin{abstract}  \selectlanguage{esperanto}
Ebeno eklumi^gas unukolore kaj tuj poste mallumi^gas, dum observanto ekmovi^gas orte el ebeno kun konstanta propra akcelo. Special-relativeco anta^udiras, ke observanto vidos luman cirklon ^ciam  en direkto mala al la ebeno, sen Dopplera efiko, kaj ke tiu cirklo ^sajnos progrese eti^gi ^gis punkto. 
%\newline \mbox{} %\newline
%\ppdu{\selectlanguage{portuguese}
%Um plano acende monocromaticamente e imediatamente ap\'os se apaga, enquanto um observador parte perpendicularmente do plano com ace\-le\-ra\c c\~ao pr\'opria constante. A relatividade especial prediz que o observador ver\'a um c\'irculo luminoso sempre na dire\c c\~ao oposta ao plano, sem efeito Doppler, e que esse c\'irculo parecer\'a encolher progressivamente at\'e um ponto.} 
%\newline \mbox{} %\newline
%\ppdu{\selectlanguage{portuguese}
%A plane lights on monochromatically and immediately after lights off, while an observer starts moving out from the plane, perpendicularly to it, with constant proper acceleration. Special relativity predicts that the observer will see a light circle always in the direction opposite to the plane, without Doppler effect, and that the circle will seem to  progressively shrink to a point.} 
\end{abstract}

\ppdu{
\begin{Parallel}[v]{\pplw}{\pprw}
%\begin{Parallel}[v]{}{}
}

\ppdu{\section*{\vspace{-2em}}\vspace{-2ex}}   %PORQUE PRECISO DISTO ?

\ppsection[0.6ex]{Enkonduko}{Introdu\c c\~ao}     %Sekcio1  %\"u 
\ppln{Supozu observanton senmovan en ebeno. Subite la tuta ebeno eligas unukoloran \mbox{lumon} en ^ciuj direktoj, dum tre mallonga \mbox{intertempo}. Sammomente la observanto ekmovi^gas orte el la ebeno, kun konstanta propra akcelo. Ni priskribas tion, kion la observanto vidas el tiu ekbrilo, la^u special-relativeco.}
\pprn{Suponha um observador parado em um plano. Subitamente o plano inteiro emite luz mono\-cro\-m\'a\-tica em todas as dire\c c\~oes, com dura\c c\~ao muito curta. No mesmo momento o observador come\c ca a se mover perpendicularmente ao plano, com ace\-le\-ra\c c\~ao pr\'opria constante. N\'os descrevemos o que o observador v\^e desse lampejo, segundo a relatividade especial.}

\ppl{Movadon de objekto kun konstanta propra akcelo studis pluraj a^utoroj: M\o ller \cite[pa^go 72]{Moller}, Rindler \cite[pa^go 49]{Rindler}, Dwayne Hamilton \cite{Hamilton}, Landau kaj Lifshitz \cite[pa^go 22]{LandauLifshitz}, Cochran \cite{Cochran}, kaj ni mem \cite{PaivaTeixeira2006, PaivaTeixeira2007a, PaivaTeixeira2007b, PaivaTeixeira2008a, PaivaTeixeira2008b}. Propra akcelo de objekto estas la akcelo mezurata per inercia referenca sistemo kie la objekto momente restas. Tiu akcelo ${\vec a}$ malegalas Newtonan akcelon ${\vec a}_N:=\dd^2{\vec x}/\dd t^2$ kaj pli ta^ugas al special-relativeco. Fakte, la rapido $v$ de objekto kun konstanta Newtona akcelo $a_N$ povas matematike superi la liman valoron de vakuo-lumo-rapido $c$, post sufi^ce da tempo. Kontra^ue, rapido de objekto kun konstanta propra akcelo $a$ neniam estos $c$.}
\ppr{O movimento de um objeto com ace\-le\-ra\c c\~ao pr\'opria constante foi estudado por v\'arios autores: M\o ller \cite[p\'ag.72]{Moller}, Rindler \cite[p\'ag.49]{Rindler}, Dwayne Hamilton \cite{Hamilton}, Landau e Lifshitz \cite[p\'ag.22]{LandauLifshitz}, Cochran \cite{Cochran}, e n\'os mesmos \cite{PaivaTeixeira2006, PaivaTeixeira2007a, PaivaTeixeira2007b, PaivaTeixeira2008a, PaivaTeixeira2008b}. ace\-le\-ra\c c\~ao pr\'opria de um objeto \'e aquela medida em um sistema inercial de refer\^encia em que o objeto esteja momentaneamente parado. Essa ace\-le\-ra\c c\~ao ${\vec a}$ difere da ace\-le\-ra\c c\~ao Newtoniana ${\vec a}_N:=\dd^2{\vec x}/\dd t^2$ e se presta melhor \`a relatividade especial. Com efeito, a velocidade $v$ de um objeto com ace\-le\-ra\c c\~ao Newtoniana $a_N$ constante pode matematicamente superar o valor limite $c$ da velocidade da luz no v\'acuo, depois de um tempo suficiente. Ao contr\'ario, a velocidade de um objeto com ace\-le\-ra\c c\~ao pr\'opria $a$ constante nunca ser\'a $c$.}
  
\ppl{Por priskribi la fenomenon, difinu $S\,':=[\,ct';z',\rho',\varphi\,'\,]$ kiel space cilindra sistemo de koordinatoj kie lumfonto estas senmova kaj okupas la tutan ebenon $z'=0$. Vidu figuron~\ref{fig.Primeira}. La observanto komence estas senmova en loko $z'\!=\!\rho'\!=\!0$, kaj ekde momento  $t'\!=\!0$ \^gi movi^gas kun konstanta propra akcelo $a$ en la pozitiva direkto de akso $z'$. ^Gia movado do estas \cite[pa^go 74]{Moller}, \cite{PaivaTeixeira2008a}}
\ppr{Para descrever o fen\^omeno, defina  $S\,':=[\,ct';z',\rho',\varphi\,'\,]$ como um sistema de coordenadas espacialmente cil\'\i ndrico em que a fonte luminosa est\'a parada e ocupa o plano $z'=0$ inteiro. Veja a figura~\ref{fig.Primeira}. O observador est\'a inicialmente parado em $z'\!=\!\rho'\!=\!0$, e a partir do momento $t'\!=\!0$ ele se move com ace\-le\-ra\c c\~ao pr\'opria $a$ constante na dire\c c\~ao positiva do eixo $z'$. Seu movimento \'e portanto  \cite[p\'ag.74]{Moller}, \cite{PaivaTeixeira2008a}}
\bea                                                   \label{ek.xyz}%01
z'=\frac{c^2}{a}\,[\,\cosh(a\tau/c)-1]\,,\hspace{3mm} \rho'=0\,,\hspace{3mm}\sinh(a\tau/c)=at'/c\,, 
\eea
\ppln{estante $\tau$ la propra intertempo de observanto ekde $t'=0$. Rimarku en lasta (\ref{ek.xyz}), ke $\tau$ egi^gas malpli rapide ol tempa koordinato $t'$ de horlo^garo de  $S\,'$. Rapido $v:=\dd z'/\dd t'$ de observanto en $S\,'$ kaj rilata faktoro $\gamma$ de tempa dilato plii^gas kiel}
\pprn{sendo $\tau$ o tempo pr\'oprio do observador decorrido desde  $t'=0$. Repare na \'ultima (\ref{ek.xyz}) que $\tau$ cresce menos rapidamente que a coordenada temporal $t'$ da cole\c c\~ao de rel\'ogios de $S\,'$. A velocidade $v:=\dd z'/\dd t'$ do observador em $S\,'$ e o correspondente fator $\gamma$ de dilata\c c\~ao temporal crescem como} 
\bea                                                    \label{ek.vg}%02
v=c\,\tanh(a\tau/c)\,,\hspace{5mm} \gamma=\cosh(a\tau/c)\,. 
\eea 

\ppl{Kelkaj interesaj rimarkoj pri akceloj \mbox{indas} mencion ^ci tie. Unue, la konstanta propra akcelo $a$ en ^ci tiu artikolo rilatas al malkonstanta Newtona akcelo $a_N(t')$ per $a=\gamma^3a_N$, estante $\gamma(t'):=1/\sqrt{1-v^2/c^2}$ kaj estante $v(t')$ la rapido de objekto. ^Car $\gamma(t')\!\geq\!1$, tial $a_N(t')\!\leq\!a$ ^ci tie.}
\ppr{Alguns reparos interessantes sobre ace\-le\-ra\c c\~oes merecem men\c c\~ao aqui. Primeiro, a ace\-le\-ra\c c\~ao pr\'opria constante $a$ neste artigo se relaciona \`a ace\-le\-ra\c c\~ao Newtoniana vari\'avel $a_N(t')$ via $a=\gamma^3a_N$, sendo  $\gamma(t'):=1/\sqrt{1-v^2/c^2}$ e sendo $v(t')$ a velocidade do objeto. Como $\gamma(t')\!\geq\!1$, ent\~ao $a_N(t')\!\leq\!a$ aqui.}

\ppl{Due, ekvacioj (\ref{ek.vg}) kaj (\ref{ek.xyz}) ka^uzas $v\approx a_Nt'$ kiam $\tau\approx0$, kvaza^u la Newtona kinematiko. Trie, la rapido $v$ apena^u proksi\-m\-i^gas al $c$, kiam $\tau\!\!\rightarrow\!\!\infty$. Tion oni komprenas: $\gamma(t')$ multe plivaloras kiam $v(t')\!\rightarrow\!c$, tial ^gi multe malfortigas la mal\-konstantan Newtonan akcelon $a_N\!=\!a/\gamma^3$.}
\ppr{Segundo, as equa\c c\~oes (\ref{ek.vg}) e (\ref{ek.xyz}) causam $v\!\approx\!a_Nt'$ quando $\tau\!\approx\!0$, como na cinem\'atica Newtoniana. Terceiro, a velocidade $v$ apenas se aproxima de $c$, quando $\tau\rightarrow\infty$. Isso se compreende: o valor de $\gamma(t')$ cresce muito quando $v(t')\rightarrow c$, assim ele torna muito fraca a ace\-le\-ra\c c\~ao Newtoniana vari\'avel $a_N=a/\gamma^3$.}   

\ppsection[0.6ex]{Percepto de lumo}{Percep\c c\~ao da luz}     %Sekcio1   \label{sek.percepto}  %\"u 
 
\begin{figure}                                                   %Figuro1
\centerline{\epsfig{file=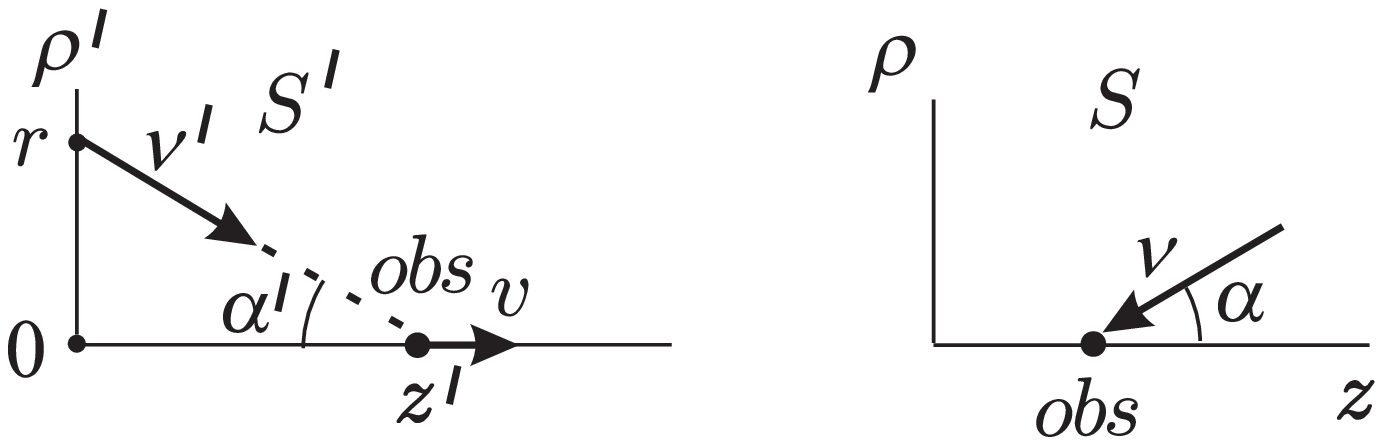,width=7cm}} 
\selectlanguage{esperanto}\caption{\selectlanguage{esperanto}En sistemo $S\,'$, lumo eligita kun frekvenco $\nu\,'$ incidas sur observanto kun rapido $v$. En sistemo $S$ de momenta senmovo de observanto, tiu lumo estas perceptata kun frekvenco $\nu$. \^Ce la kinematikaj kondi\^coj supozataj en la teksto, special-relativeco anta^uvidas egala^jojn $\nu\!=\!\nu\,'$ kaj $\alpha=\alpha'$. 
\ppdu{\newline \selectlanguage{portuguese}Figura~1: No sistema $S\,'$, luz emitida com freq\"u\^encia $\nu\,'$ incide sobre um observador que se move com velocidade $v$. No sistema $S$ de imobilidade moment\^anea do observador, essa luz \'e percebida com freq\"u\^encia $\nu$. Nas condi\c c\~oes cinem\'aticas supostas no texto, a relatividade especial prev\^e as igualdades $\nu\!=\!\nu\,'$ e $\alpha=\alpha'$.}}  
\label{fig.Primeira} 
\end{figure}  

\ppln{Por studi percepton de ekbrilo de ebeno esti^gu, en sistemo $S\,'$, lumo eligita el cirklo $[z',\rho',\varphi\,']=[0,r,\forall]$ en momento $t'=0$, kun onda 4-vektoro (vidu figuron~\ref{fig.Primeira})}
\pprn{Para estudar a percep\c c\~ao do lampejo do plano seja, no sistema $S\,'$, a luz emitida do c\'\i rculo  $[z',\rho',\varphi\,']=[0,r,\forall]$ no momento $t'=0$, com 4-vetor de onda (veja a figura~\ref{fig.Primeira})}
\bea                                              \label{ek.kmulinha}%03
k^{\mu'}:=\nu\,'\,[\,1;\,\cos\alpha'\,,-\sin\alpha', 0\,]\ . 
\eea
\ppln{Tiu lumo incidas sur la observanto  en $(t',z')$ se $ct'=\sqrt{{z'}^2+r^2}$. Uzante $z'(\tau)$ kaj $t'(\tau)$ el (\ref{ek.xyz}), kaj uzante $\cos\alpha'=z'/(ct')$ el figuro~\ref{fig.Primeira}, rezulti^gas}
\pprn{Essa luz incide sobre o observador em $(t',z')$ se $ct'=\sqrt{{z'}^2+r^2}$. Usando $z'(\tau)$ e $t'(\tau)$ da (\ref{ek.xyz}), e usando $\cos\alpha'=z'/(ct')$ da figura~\ref{fig.Primeira}, resulta em}  
\bea                                               \label{ek.cosalfa}%04
r=\frac{2c^2}{a}\sinh(a\tau/2c)\,, \hspace{2mm} \cos\alpha'=\tanh(a\tau/2c)\,. 
\eea
\ppln{La unua esprimo indikas tiun cirklon de la ebeno, kiun vidas la observanto en propratempo $\tau$. ^Car $\dd r/\dd\tau\approx c$ kiam $\tau\approx0$, tial la radiuso $r$ de luma cirklo komence egi^gas kun rapido $c$. Poste $r$ egi^gas ankora^u pli rapide. Pro la dua esprimo, la angulo $\alpha'$ en figuro~\ref{fig.Primeira} eti^gas ekde  $\pi/2$ ^gis 0, dum la propratempo $\tau$ kaj la loko $z'$ de observanto egi^gas amba^u de 0 ^gis $\infty$.}
\pprn{A primeira express\~ao indica o c\'\i rculo do plano que o observador v\^e no tempo pr\'oprio $\tau$. Como $\dd r/\dd\tau\approx c$ quando $\tau\approx0$, ent\~ao o raio $r$ do c\'\i rculo luminoso inicialmente cresce com velocidade $c$. Depois $r$ cresce ainda mais rapidamente. Pela segunda express\~ao, o \^angulo $\alpha'$ na figura~\ref{fig.Primeira} diminui desde $\pi/2$ at\'e 0, enquanto o tempo pr\'oprio $\tau$ e a posi\c c\~ao $z'$ do observador crescem ambos de 0 at\'e $\infty$.} 

\ppl{Por priskribi luman percepton, la observanto kun rapido $v$ uzas inercian sistemon kie ^gi momente estas senmova, en momento de percepto. Do ni elektas sistemon $S$ kun rapido $v$ rilate al $S\,'$, kaj havante akson $z$ sur akso $z'$, kiel en figuro~\ref{fig.Primeira}.}
\ppr{Para descrever a percep\c c\~ao da luz, o observador com velocidade  $v$ usa um sistema inercial em que ele est\'a momentaneamente parado, no momento da percep\c c\~ao. Ent\~ao n\'os escolhemos um sistema $S$ com velocidade $v$ relativamente a $S\,'$, e tendo eixo $z$ sobre o eixo $z'$, como na figura~\ref{fig.Primeira}.} 

\ppl{En sistemo $S$, komponantoj de onda 4-vektoro de lumo eligita el $r$ estas kalkulitaj el komponantoj (\ref{ek.kmulinha}) per Lorentzaj transformoj, kiel}
\ppr{No sistema $S$, as componentes do 4-vetor de onda da luz emitida de  $r$ s\~ao calculadas das componentes (\ref{ek.kmulinha}) por transforma\c c\~oes de Lorentz, como}  
\bea                                                   \label{ek.kmu}%05
\nu\!=\!\gamma \nu\,'[1-(v/c)\cos\alpha'],\hspace{2mm} k^{z}\!\!=\!\gamma \nu\,'(\cos\alpha'-v/c),\hspace{2mm} k^{\rho}\!\!=\!-\nu\,'\sin\alpha', \hspace{2mm} k^{\varphi}\!\!=\!0 \ .
\eea 
\ppln{^Car la observanto foriras de fonto, oni naive povus esperi, ke frekvenco $\nu$ de perceptata lumo estus plieta ol frekvenco $\nu\,'$ de eligita lumo. Tamen tio ne veri^gas, kiel ni konstatas uzante la unuan el (\ref{ek.kmulinha}), kaj poste (\ref{ek.vg}) kaj (\ref{ek.cosalfa}):}
\pprn{Como o observador se afasta da fonte, poder-se-ia ingenuamente  esperar que a freq\"u\^encia $\nu$ da luz percebida fosse menor que a freq\"u\^encia $\nu\,'$ da luz emitida. Entretanto isso n\~ao \'e verdade, como n\'os constatamos usando a primeira das (\ref{ek.kmulinha}), e depois (\ref{ek.vg}) e (\ref{ek.cosalfa}):} 
\bea                                                \label{ek.iguais}%06
\frac{\nu}{\nu\,'}=\gamma\left[\,1-\frac{v}{c}\cos\alpha'\right]=\cosh\left(\frac{a\tau}{c}\right)\left[\,1-\tanh\left(\frac{a\tau}{c}\right)\tanh\left(\frac{a\tau}{2c}\right)\right]\equiv 1\,. 
\eea
\ppln{Do la Dopplera faktoro $D:=\nu/\nu\,'$ estas~1, t.e., ne estas Dopplera efiko. Tiu fakto ne estus esperata, precipe ^car ^gi okazas en ^ciu momento $\tau$, kaj por iu ajn valoro de konstanta propra akcelo $a$.}
\pprn{Portanto o fator Doppler $D:=\nu/\nu\,'$ \'e~1, i.e., n\~ao h\'a efeito Doppler. Esse fato n\~ao seria esperado, principalmente porque ele ocorre em todos os momentos $\tau$, e para qualquer valor da ace\-le\-ra\-\c c\~ao pr\'opria constante $a$.} 

\ppl{Alia malesperata fakto estas, ke malkonstantaj komponantoj $k^{z}$ kaj $k^{z'}$ ^ciam havas kontra^uajn signumojn, kaj saman modulon. Tion oni konstatas uzante $k^{z}$ el (\ref{ek.kmu}) kaj $k^{z'}$ el (\ref{ek.kmulinha}), poste (\ref{ek.vg}) kaj (\ref{ek.cosalfa}):}
\ppr{Outro fato inesperado \'e que as componentes vari\'aveis  $k^{z}$ e $k^{z'}$ sempre t\^em sinais opostos, e mesmo m\'odulo. Isso se constata usando $k^{z}$ da (\ref{ek.kmu}) e $k^{z'}$ da (\ref{ek.kmulinha}), depois (\ref{ek.vg}) e (\ref{ek.cosalfa}):}
\bea                                                \label{ek.opostos}%07
\frac{k^{z}}{k^{z'}}=\frac{\gamma[\,\cos\alpha'-v/c]}{\cos\alpha'}=\frac{\cosh(a\tau/c)[\,\tanh(a\tau/2c)-\tanh(a\tau/c)]}{\tanh(a\tau/2c)}\equiv -1\,. 
\eea
\ppln{^Car anka^u okazas $k^{\rho}\!=\!k^{\rho'}$, tial la observanto en $S$ vidas la fonton anta^uen, kun angulo $\alpha$ sama al la angulo $\alpha'$ ^ce $S'$, kiel montras figuro~\ref{fig.Primeira}. El (\ref{ek.cosalfa}), $\alpha'$ eti^gas dum la movado, tial anka^u $\alpha$ eti^gas dum la movado, kaj do la observanto vidas la luman cirklon progrese nuli^gi.}
\pprn{Como ocorre tamb\'em $k^{\rho}\!=\!k^{\rho'}$, ent\~ao o observador em $S$ v\^e a fonte \`a sua dianteira, com \^angulo $\alpha$ igual ao \^angulo $\alpha'$ em $S'$, como a figura~\ref{fig.Primeira} mostra. Em (\ref{ek.cosalfa}), $\alpha'$ diminui durante o movimento, da\'\i\,\, tamb\'em $\alpha$ diminui durante o movimento, e portanto o observador v\^e o c\'\i rculo luminoso progressivamente se anular.}

\ppl{Ripetinde, la observanto vidas la lumon nur se li estas dorse al eliganta ebeno. Nomi^gas aberacio, iu ajn ^san^go de ^sajna direkto de objekto pro movado de observanto. Aberacion de objekto ni nomu {\em forta}, se la objekto estas malanta^u la movi^ganta observanto en sistemo $S\,'$ sed estas vidata anta^uen per observanto en sistemo $S$. El~(\ref{ek.opostos}), forta aberacio okazas se $v>c\,\cos\alpha'$, t.e., se la rapido de observanto estas pligranda ol la projekcio de rapido de lumo sur vojo de observanto (vidu figuron~\ref{fig.Primeira}).} 
\ppr{Vale repetir, o observador v\^e a luz somente se ele est\'a de costas para o plano emissor. Chama-se aberra\c c\~ao, qualquer mudan\c ca de dire\c c\~ao aparente de um objeto devido \`a movimenta\c c\~ao do observador. Vamos chamar de {\em forte} a aberra\c c\~ao, se o objeto est\'a atr\'as do observador em movimento no sistema $S\,'$ mas \'e visto de frente pelo observador no sistema $S$. Pela~(\ref{ek.opostos}), a aberra\c c\~ao forte ocorre se $v>c\,\cos\alpha'$, i.e., se a velocidade do observador for maior que a proje\c c\~ao da velocidade da luz sobre o caminho do observador (veja a figura~\ref{fig.Primeira}).}

\ppl{Speciale, fortan aberacion de objekto oni nomu {\em simetria} se anguloj $\alpha$ kaj $\alpha'$ estas samaj. Pro (\ref{ek.vg}) kaj (\ref{ek.cosalfa}), ^ci tie fari^gas simetria aberacio de ^ciu vidata cirklo. Estas mirinda, ke tiu fenomeno okazas ekde komenco de movado de observanto, e^c kun iu ajn valoro de konstanta propra akcelo.}
\ppr{Em particular, chame-se {\em sim\'etrica} a aberra\-\c c\~ao forte se os \^angulos  $\alpha$ e $\alpha'$ forem iguais. Devido a (\ref{ek.vg}) e (\ref{ek.cosalfa}), aqui ocorre aberra\c c\~ao sim\'etrica de todo c\'\i rculo visto. \'E admir\'avel que esse fen\^omeno ocorra desde o come\c co do movimento do observador, e com qualquer valor da ace\-le\-ra\-\c c\~ao pr\'opria constante.}

\ppsection[0.6ex]{Konkludo}{Conclus\~ao}       %Sekcio3  %\"u

\ppln{Ni montris ke ne okazas Dopplera efiko sed estas simetria forta aberacio, tiel ke la lumaj cirkloj estas simetrie vidataj en la direkto mala al la ekbrilanta ebeno. Komence la observanto vidas malgrandan cirklon ^ce $\alpha=90^o$, poste la angulo progrese eti^gas ^gis nuli.}
\pprn{N\'os mostr\'amos que n\~ao ocorre efeito Doppler mas ocorre aberra\c c\~ao forte sim\'etrica, tal que os c\'\i rculos luminosos s\~ao vistos simetricamente na dire\c c\~ao oposta ao plano brilhante. Inicialmente o observador v\^e um c\'\i rculo pequeno em $\alpha=90^o$, depois o \^angulo progressivamente diminui at\'e se anular.} 

\ppl{La metodo, kiun ni uzis ^ci tie por kalkuli Doppleran efikon, estas tre kompakta. Tamen la metodo uzata en niaj anta^uaj artikoloj~\cite{PaivaTeixeira2006,PaivaTeixeira2007a,PaivaTeixeira2007b,PaivaTeixeira2008a,PaivaTeixeira2008b} malka^sas gravan proprecon de propra akcelo.}
\ppr{O m\'etodo que n\'os us\'amos aqui para calcular o efeito Doppler \'e bastante compacto. Entretanto o m\'etodo usado em nossos artigos anteriores~\cite{PaivaTeixeira2006,PaivaTeixeira2007a,PaivaTeixeira2007b,PaivaTeixeira2008a,PaivaTeixeira2008b} re\-ve\-la uma propriedade importante da acelera\c c\~ao pr\'opria.}

\ppl{Tion celante, konsideru en sistemo $S\,'$ de figuro~\ref{fig.Primeira} lumsignalon eligitan en momento $t_f'$ kaj perceptatan en momento $t_o'$. Do la elig-eniga ekvacio estas $c(t_o'-t_f')=\sqrt{{z'}^2(t_o')+r^2}$; ^ci tie $z'(t_o')$ estas kiel en (\ref{ek.xyz}), kaj $r$ estas radiuso de cirklo vidata per observanto en momento $t_o'$. Diferenciante kaj poste farante $t_f'=0$ kaj $\dd t_f'=\dd\tau_f$, ni ricevas $\dd\tau_f=\dd\tau_o$, kaj do $D=\dd\tau_f/\dd\tau_o=1$. Tiu rezulto aperas ka^sita en anta^ua artikolo~\cite{PaivaTeixeira2008a}. Fakte, ^ciu lumsignalo en nuna artikolo estas kiel la signalo en limo inter fazoj~1 kaj 2 en sekcio~5 de~\cite{PaivaTeixeira2008a}.}
\ppr{Com isso em vista, considere no sistema $S\,'$ da figura~\ref{fig.Primeira} um sinal de luz emitido no momento $t_f'$ e percebido no momento $t_o'$. Ent\~ao a equa\c c\~ao de emiss\~ao-recep\c c\~ao \'e $c(t_o'-t_f')=\sqrt{{z'}^2(t_o')+r^2}$; aqui $z'(t_o')$ \'e como em (\ref{ek.xyz}), e $r$ \'e o raio do c\'\i rculo visto pelo observador no momento $t_o'$. Diferenciando e depois fazendo $t_f'=0$ e $\dd t_f'=\dd\tau_f$, n\'os recebemos $\dd\tau_f=\dd\tau_o$, e portanto $D=\dd\tau_f/\dd\tau_o=1$. Esse resultado aparece escondido em um artigo anterior~\cite{PaivaTeixeira2008a}. Com efeito, todo sinal luminoso no atual artigo \'e como o sinal no limite entre as fases~1 e 2 na se\c c\~ao~5 de~\cite{PaivaTeixeira2008a}.}

\ppl{Simile oni povus rericevi ^ciujn ajn rezultojn el anta^uaj sekcioj en ^ci tiu artikolo. Sed nun interesas nur la kompreno de graveco de konstanteco de propra akcelo al nuligo de Dopplera efiko.}
\ppr{Semelhantemente se poderia reobter todos os resultados das se\c c\~oes anteriores neste artigo. Por\'em agora interessa somente a compreens\~ao da import\^ancia da const\^ancia da ace\-le\-ra\c c\~ao pr\'opria para a anula\c c\~ao do efeito Doppler.}

\ppl{Do, konsideru lumsignalon eligitan je $t_f'=0$, kaj inercian sistemon de momenta senmovo de observanto je tiu momento. En tiu sistemo, observanto havas konstantan propran akcelon $a$ en direkto $z'$, kaj fonto restas en la sama $z'=0$ kaj distancas $r$ perpendikulare.}
\ppr{Considere ent\~ao um sinal luminoso emitido em $t_f'=0$, e um sistema inercial de imobilidade moment\^anea do observador naquele momento. Nesse sistema o observador tem ace\-le\-ra\-\c c\~ao pr\'opria constante $a$ na dire\c c\~ao $z'$, e a fonte est\'a parada no mesmo $z'=0$ e dista $r$ na perpendicular.}

\ppl{Plu, konsideru duan signalon eligitan je momento $\dd t_f'$, kaj inercian sistemon de momenta senmovo de observanto je tiu momento. En tiu dua sistemo, observanto havas konstantan propran akcelon $a$ en direkto $z'$, estas en loko $\dd z'$, kaj la fonto havas infiniteziman rapidon en direkto malpozitiva de $z'$ kaj distancas $r$ perpendikulare. ^Car la interspaco trakurata per la observanto estas duagrada en $\dd t_f'$, tial la observanto estas konsiderata ankora^u en la komenca loko.}
\ppr{Considere ainda um segundo sinal, emitido no momento $\dd t_f'$, e um sistema inercial de imobilidade moment\^anea do observador naquele momento. Nesse segundo sistema o observador tem ace\-le\-ra\c c\~ao pr\'opria constante $a$ na dire\c c\~ao $z'$, est\'a na posi\c c\~ao $\dd z'$, e a fonte tem uma velocidade infinitesimal na dire\c c\~ao negativa de $z'$ e dista $r$ perpendicularmente. Como a dist\^ancia percorrida pelo observador \'e quadr\'atica em $\dd t_f'$, ent\~ao o observador \'e considerado ainda na posi\c c\~ao inicial.}

\ppl{Por movado de lumsignalo, la rapido de fonto ne gravas. Anka^u rimarku, ke la infinitezima rapido de la fonto kontribuas duagrade al $\gamma$, do $\dd\tau_f=\dd t_f'$. Tial la intertempo inter eligo kaj percepto estas la sama por la du signaloj, t.e., $\tau_o-0=(\tau_o +\dd\tau_o)-(0+\dd\tau_f)$.  Do $\dd\tau_o=\dd\tau_f$ kaj do $D:=\dd\tau_f/\dd\tau_o=1$, t.e., Dopplera efiko ne estas. Rimarku, ke tiu argumento nur validas kun propra akcelo, ^car ^ci tiu estas difinata en inerciaj sistemoj de momenta senmovo.}
\ppr{Para o movimento do sinal luminoso, a velocidade da fonte n\~ao importa. Repare tamb\'em que a velocidade infinitesimal da fonte contribui quadraticamente ao $\gamma$, portanto $\dd\tau_f=\dd t_f'$. Ent\~ao o intervalo de tempo entre a emiss\~ao e a percep\c c\~ao \'e o mesmo para os dois sinais, i.e., $\tau_o-0=(\tau_o +\dd\tau_o)-(0+\dd\tau_f)$.  Da\'\i\,\, $\dd\tau_o=\dd\tau_f$ e portanto $D:=\dd\tau_f/\dd\tau_o=1$, i.e., n\~ao h\'a efeito Doppler. Repare que esse argumento vale somente com ace\-le\-ra\c c\~ao pr\'opria, porque esta \'e definida em sistemas inerciais de imobilidade moment\^anea.}

\ppl{Ni opinias, ke la rezultoj de ^ci tiu artikolo evidentigas la mirindan gravecon de konstanta propra akcelo ^ce special-relativeco.}
\ppr{N\'os achamos que os resultados deste artigo evidenciam a admir\'avel import\^ancia da ace\-le\-ra\c c\~ao pr\'opria constante na relatividade especial.}

\ppdu{\end{Parallel}} % \"
\end{document}